\documentclass[twocolumn,prb,aps,superscriptaddress]{revtex4-1}

\usepackage{graphicx}
\usepackage{amsmath}
\usepackage{bm}

\newcommand{\D}{\mathrm{d}}

\newcommand{\lsco}{La$_{2-x}$Sr$_x$CuO$_4$}

\begin{document}

\title{Optical conductivity of  overdoped cuprate superconductors: application to LSCO}

\author{N.~R.~Lee-Hone}
\affiliation{Department of Physics, Simon Fraser University, Burnaby, BC, V5A~1S6, Canada}
\author{V. Mishra}
\affiliation{Computer Science and Mathematics Division, Oak Ridge National Laboratory, Oak Ridge, TN 37831, USA}
\author{D.~M.~Broun}
\affiliation{Department of Physics, Simon Fraser University, Burnaby, BC, V5A~1S6, Canada}
\affiliation{Canadian Institute for Advanced Research, Toronto, ON, MG5 1Z8, Canada}

\author{P.~J. Hirschfeld}
\affiliation{Department of Physics, U. Florida, Gainesville FL 32611}

\begin{abstract}
We argue that recent  measurements on both the superfluid density and the optical conductivity of high-quality \lsco\ films can be understood almost entirely within the theory of disordered BCS $d$-wave superconductors.  
The large scattering rates deduced from experiments are shown to arise predominantly from weak scatterers, probably the Sr dopants out of the CuO$_2$ plane, and correspond to significant suppression of $T_c$ relative to a pure reference state with the same doping.  
Our results confirm the ``conventional'' viewpoint that the overdoped side of the cuprate phase diagram can be viewed as approaching the BCS weak-coupling description of the superconducting state, with significant many-body renormalization of the plasma frequency. They suggest that, while some of the decrease in $T_c$ with overdoping may be due to weakening of the pairing, disorder plays an essential role.

\end{abstract}

\maketitle{}

\section{Introduction}

The cuprate phase diagram has been the subject of considerable controversy over the 30 years since the discovery of high-$T_c$ superconductivity.\cite{Keimeretal:2014}  Of the various exotic phases observed, including pseudogap, charge order, etc., most are located on the underdoped side.  The \mbox{$d$-wave} superconducting phase is thought to be the simplest to understand, particularly on the overdoped side, where in many systems it exists without obvious competing or coexisting orders of other types.     Recently an experiment cast doubt on this simple picture of a ``garden variety'' BCS $d$-wave superconductor.  Bo{\v z}ovi{\'c} et al.\cite{Bozovic:2016ei} measured the  superfluid density of a finely spaced set of high-quality epitaxial grown films of overdoped \lsco,  and showed  that the superfluid density $\rho_s$ and the superconducting transition temperature $T_c$ approached zero together as a function of doping. This by itself contradicts BCS theory, which predicts that the $T=0$ superfluid density should be simply the carrier density (in appropriate units) independent of $T_c$ in a clean system.  In a dirty superconductor, the superfluid density correlates with $T_c$,\cite{MaLee:1985,SmithAmbegaokar:1992,Puchkaryov1998,Kogan:2013eh} but  the $T$-dependence of the penetration depth deduced from the measurements in Ref.~\onlinecite{Bozovic:2016ei} is nearly linear down to the lowest temperatures, suggesting that the films are in fact largely free of disorder.  Under the assumption that the systems are clean,  Bo{\v z}ovi{\'c} et al.\cite{Bozovic:2016ei} concluded that the scaling of $\rho_s$ with $T_c$ implies a substantial reduction in superfluid density relative to the nominal carrier density, inconsistent with a BCS description.

Subsequently, two of the present authors, with J.~S.~Dodge, analyzed the data of Ref.~\onlinecite{Bozovic:2016ei} and reached rather  different conclusions.\cite{Lee-Hone:2017}  They pointed out that if the Sr dopants  were treated as weak  (Born limit) scatterers, the lack of a $T^2$ term in the penetration depth down to the lowest measurement temperatures of Bo{\v z}ovi{\'c} et al.\ could easily be understood.  It has been known for many years that in this limit the quasiparticle states near the $d$-wave nodes are broadened by disorder, and that this broadening occurs significantly over an energy range that is exponentially small in $\Gamma_N/\Delta_0$, where $\Gamma_N$ is the normal state disorder scattering rate and $\Delta_0$ the $d$-wave gap maximum.  For clean systems, states at energies greater than this scale are largely unaffected, and the penetration depth retains its linear dependence. Even for systems where $\Gamma_N/\Delta_0$ becomes appreciable, however, the linear-$T$ behavior obtains over a surprisingly large range.\cite{Kogan:2013kf,Lee-Hone:2017}  Band structure effects can also enhance this quasi-linearity in the same intermediate temperature range.  It was shown in Ref.~\onlinecite{Lee-Hone:2017} that the data could be fit extremely well over the entire range of doping and temperature using a single choice of disorder parameters and the known doping-dependent Fermi surface measured by ARPES.\cite{Yoshida:2006hw}  This comparison is summarized in Fig.~\ref{fig1}.

The most logical way to distinguish between the disorder scenario of Ref.~\onlinecite{Lee-Hone:2017} and more exotic explanations is to measure the spectral weight of the condensed and uncondensed carriers directly using optical probes.  Recently, Mahmood et al.\ have performed terahertz (THz) spectroscopy\cite{Bilbro:2011jj,Mahmood:2017} on  films very similar to those used in the Bo{\v z}ovi{\'c} et al.\ superfluid density measurements.\cite{Bozovic:2016ei}  As anticipated, they found that  a significant fraction of
the carriers remained uncondensed in a broad Drude-like
peak at low temperatures. They showed consistency with the earlier superfluid density measurements, but argued that the broad uncondensed
spectrum implied a quasiparticle scattering rate too large  to be consistent with the linear-$T$-penetration depth observed.  Mahmood et al.\ therefore concluded that their results definitively ruled out a disorder-based explanation of the overdoped data.

In this work, we re-examine the conductivity of a \mbox{$d$-wave} superconductor in the presence of disorder with a view towards refining the interpretation of Mahmood et al.'s work.  We argue that the weak scattering limit used predominantly in Ref.~\onlinecite{Lee-Hone:2017} yields a self-energy that grows roughly with energy up to a scale of $\Delta_0$ in the clean limit, becoming roughly constant when the scattering rate is a significant fraction of $T_{c0}$, the critical temperature of the pure system, but before superconductivity is fully suppressed.    This implies that the conductivity spectrum has a Drude-like shape even in the superconducting state, for reasonably dirty systems.  This  result is qualitatively different from earlier works on the optical properties of $d$-wave superconductors, which focused primarily on underdoped to optimally doped systems where strong in-plane defects dominated the scattering.  Using {\it exactly the same model and parameters} employed in Ref.~\onlinecite{Lee-Hone:2017} to describe superfluid density,  we obtain a qualitatively very reasonable fit to the results of Mahmood et al.  Nearly perfect fits can be obtained with slight further fine tuning.  While for each sample, substantial pairbreaking is indeed involved, the Born limit scattering rate parameter $\Gamma_N/T_{c0}$ is still sufficiently small compared to 1 so as to allow the near-linear $T$ dependence of the penetration depth in this limit. 

Our analysis thus implies that a BCS-like disordered $d$-wave scenario can indeed explain the unusual dependence of the superfluid density on $T_c$ that is observed, and has further implications for the phase diagram as well. Taken at face value, our analysis suggests that  the superconducting dome is suppressed at high doping in  part due to disorder, but also indicates the existence of a reference ``pure'' $T_{c0}$ that is significantly higher than the observed $T_c$, implying that such critical temperatures might be achieved if the given doping level could be reached without potential scattering.    We discuss how $T_{c0}$ might depend on interactions, as well as possible roles of spin fluctuation scattering and forward scattering due to out-of-plane defects. 

\section{Theory}

\label{theory}

\subsection{Dirty d-wave superconductivity}
The so-called ``dirty $d$-wave" theory of cuprate superconductivity is a simple extension of the original field-theoretical formulation of the theory of disordered superconductors by Abrikosov and Gor'kov.\cite{AAAbrikosov:1960}  This theory was applied for the first time to unconventional (nodal) superconductors  independently by Gor'kov and Kalugin\cite{LPGorkov:1985} and by Rice and Ueda,\cite{KUeda:1985} and extended to arbitrary impurity phase shifts by Hirschfeld et al.\cite{Hirschfeld:1986ii} and Schmitt-Rink et al.\cite{Schmitt-Rink:1986}  Impurity effects on the superfluid density of unconventional superconductors were treated within the same formalism by Gross et al.\cite{Gross:1986} for $p$-wave superconductors, and applied to $d$-wave systems by Prohammer and Carbotte,\cite{PROHAMMER:1991p557} and Hirschfeld and Goldenfeld.\cite{Hirschfeld:1993cka}
The latter work focused on strong scattering, in an attempt to explain Zn-substituted YBCO penetration depth experiments by Bonn, Hardy and co-workers that had made the case for $d$-wave superconductivity in cuprates.\cite{HARDY:1993p632,Achkir:1993}  In the unitarity limit appropriate for the strong Zn scatterer in the CuO$_2$ plane, the linear density of states $N(\omega)\sim \omega$ in the pure $d$-wave superconductor was found to give way to a constant residual $N(0) \sim \gamma$, leading to a $T^2$ term in the penetration depth over a range of energies roughly equal to $\gamma$. Above this range, in the unitarity limit, states are relatively unaffected by disorder.\cite{Hirschfeld:1993cka}  The success of this early work has led to the impression that disorder always gives rise to immediate, strong asymptotic $T^2$ dependence in the penetration depth at low temperatures.  That this is not the case in the weak scattering (Born) limit, at least in practical terms  was noted by Hirschfeld et al.,\cite{Hirschfeld:1994} and recently reemphasized by Kogan et al.\cite{Kogan:2013kf} and in a work by two of the current authors.\cite{Lee-Hone:2017}

\begin{figure}[t]			
\includegraphics[width=1\columnwidth]{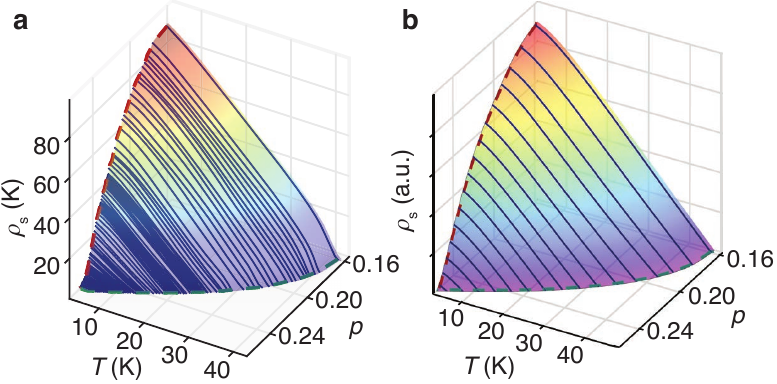}
\caption{Comparison of superfluid density data (a) measured on epitaxially grown \lsco\ thin films by Bo{\v z}ovi{\'c} et al.\cite{Bozovic:2016ei} with the (b) disorder-based theory of Lee-Hone et al.\cite{Lee-Hone:2017} \mbox{$p$ is hole doping}. Impurity parameters from that paper, used to make the curves in (b), are identical to one of the scenarios used in the current paper to study the optical conductivity.  Note both experimental data and theory are plotted only down to $T_\mathrm{min}\sim 3$~K; this obscures slight curvature below this scale in the theory. }
\label{fig1}
\end{figure}

We begin with the Nambu space Green's function for a dirty $d$-wave superconductor, written as 
\begin{eqnarray}
G(\mathbf{k},i\omega_n)=-\frac{i\tilde\omega_n \tau_0 + \Delta_{\bf k} \tau_1 + \xi_{\bf k} \tau_3} {\tilde\omega^2_n + \Delta_{\bf k}^2 + \xi_{\bf k}^2}\;,
\end{eqnarray}
where $\Delta_\mathbf{k}$ is the $d$-wave superconducting gap, $\xi_{\bf k}$ is the single-particle dispersion relative to the Fermi level, $\tau_i$ are the Pauli matrices, and $\tilde \omega_n$ is a renormalized Matsubara frequency that, in the self-consistent $t$-matrix approximation (SCTMA),\cite{Hirschfeld:1986ii, Schmitt-Rink:1986} follows 
\begin{align}
\tilde \omega_n & \equiv \tilde \omega(\omega_n)  = \omega_n + i \Sigma\\
& =\omega_n + \pi \Gamma \frac{\langle N_\mathbf{k}(\tilde \omega_n) \rangle_\mathrm{FS}}{c^2 + \langle N_\mathbf{k}(\tilde \omega_n) \rangle_\mathrm{FS}^2}\;.
\label{tmatrix}
\end{align}
Here the particular form of the self energy $\Sigma$ is
for a single type of scatterer characterized by parameters $(\Gamma,c)$, where \mbox{$c$ is the cotangent} of the scattering phase shift, $\Gamma$ is a scattering parameter proportional to the concentration of impurities, and
\begin{equation}
N_\mathbf{k}(\tilde \omega_n) = \frac{\tilde \omega_n}{\sqrt{\tilde \omega_n^2 + \Delta_\mathbf{k}^2}}\;.
\end{equation}
We have used the definition  $\langle ... \rangle_\mathrm{FS}$ as a Fermi surface angular average defined by
\begin{equation}
\langle ...\rangle_\mathrm{FS} \equiv \frac{1}{N_0}\int_0^{2 \pi} N_\phi (...) \mathrm{d} \phi\;,
\end{equation}
where 
\begin{equation}
N_\phi = \frac{1}{2 \pi^2 \hbar d} \frac{|k_F|^2}{\mathbf{k}_F \cdot \mathbf{v}_F}
\end{equation}
is the angle-resolved density of states on the Fermi surface, $N_0 = \int  N_\phi \mathrm{d} \phi$ is the integrated density of states (including both spin channels), and $d$ is the spacing of the two-dimensional conducting layers, in \lsco\ taking the value \mbox{$d = 13.15/2 = 6.57$~\AA}.  The Fermi wavevector,  $\mathbf{k}_F$, and Fermi velocity, $\mathbf{v}_F$, are both functions of $\phi$, the momentum angle on the Fermi surface.

In much of this paper we will consider a specific model that was found in Ref.~\onlinecite{Lee-Hone:2017} to explain the superfluid density results well.  It is assumed that scattering is determined by a large concentration of Born scatterers ($c \gg 1$, presumably the out-of-plane Sr dopants), and a small concentration of unitarity scatterers ($c=0$, possibly Cu vacancies).  For this model the self-consistency condition reads
\begin{eqnarray}
\tilde{\omega}_n &=& \omega_n + \Gamma_N^{B} \langle N_{\bf k}(\tilde \omega_n) \rangle_\mathrm{FS} + \frac{\Gamma_N^U}{\langle N_{\bf k}(\tilde \omega_n) \rangle_\mathrm{FS}}\;,
\label{unitarityBorn}
\end{eqnarray}
with $\Gamma_N^U/\pi = 1$~K and $\Gamma_N^B/\pi =17$~K.

Note that for a $d$-wave order parameter, which averages to zero over the Fermi surface, there is no  renormalization of the gap $\Delta_\mathbf{k}$ by nonmagnetic pointlike scatterers.  Impurities are pairbreaking, however, and suppress the gap through the effect of disorder on $\tilde \omega_n$ in the gap equation
\begin{equation}
\Delta_\mathbf{k} =2 \pi T N_0 \sum_{\omega_n > 0}^{\omega_0} \left\langle V_{\mathbf{k},\mathbf{k}^\prime} \frac{\Delta_{\mathbf{k}^\prime}}{\sqrt{\tilde \omega_n^2 + \Delta_{\mathbf{k}^\prime}^2}}\right\rangle_\mathrm{FS}\;,
\end{equation}
where $\Delta_\mathbf{k}$ is the gap parameter at wave-vector $\mathbf{k}$, \mbox{$\omega_n = 2 \pi T (n + \tfrac{1}{2})$} are the fermionic  Matsubara frequencies, $V_{\mathbf{k},\mathbf{k^\prime}}$ is the  pairing interaction, and $\omega_0$ is a high energy cutoff.
This equation as $\Delta_{\bf k}\rightarrow 0$ determines the critical temperature $T_c$, which is suppressed according to the universal Abrikosov--Gor'kov formula\cite{AAAbrikosov:1960} with pairbreaking parameter twice the normal state single-particle scattering rate, $\Gamma_N=\Gamma/(1+c^2)$.

\begin{figure*}[t]			\includegraphics[width=1\textwidth]{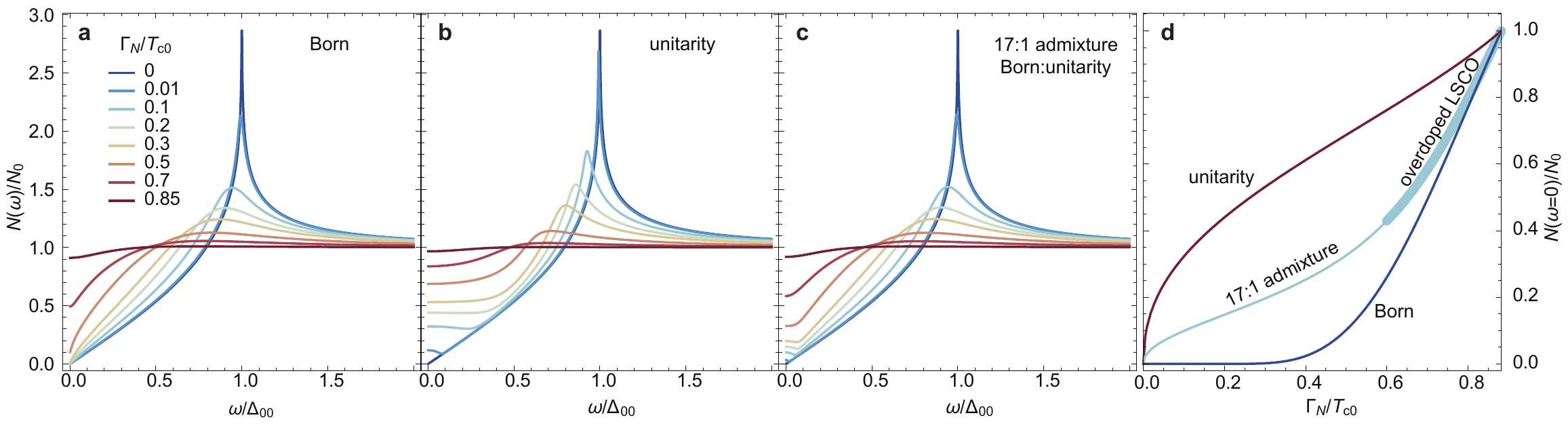}
			\caption{Predictions of dirty $d$-wave theory for the density of states $N(\omega)$ vs.\ $\omega/\Delta_{00}$, where $\Delta_{00}$ is the clean-limit, zero-temperature gap magnitude. (a) Born limit scatterers, for various normal state relaxation rates $\Gamma_N$; (b) unitarity limit scatterers; and (c) the mixture of Born and unitarity scatterers  discussed in Ref.~\onlinecite{Lee-Hone:2017}. (d) Variation of residual density of states with $\Gamma_N$, proportional to impurity concentration, for the three impurity cases.}
\label{fig2}
\end{figure*}

\begin{figure*}[htb]			\includegraphics[width=1\textwidth]{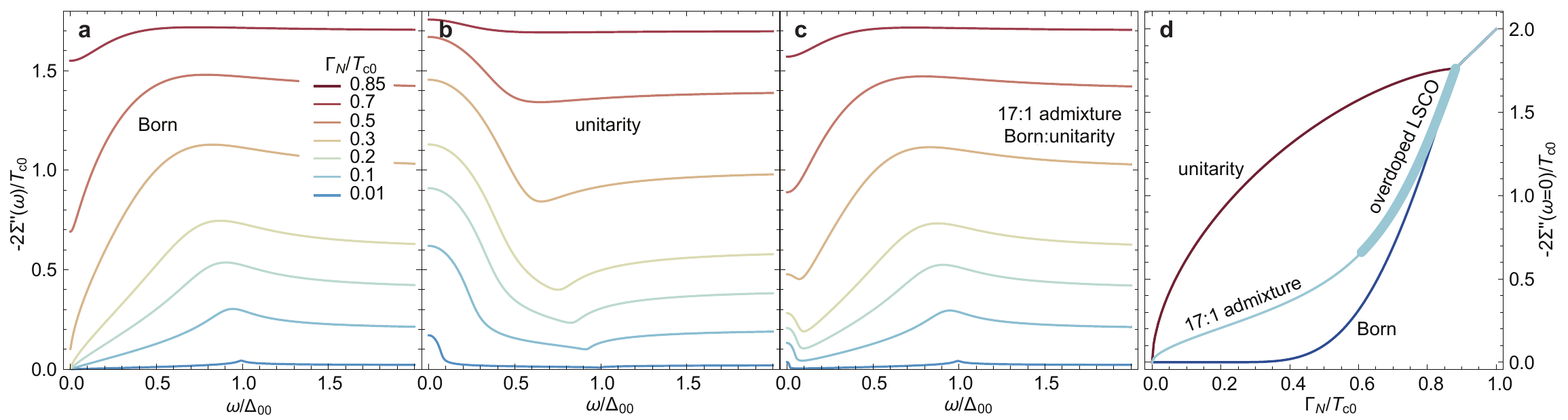}
			\caption{Superconducting state scattering rate -2Im$\Sigma(\omega)$ vs.\ $\omega/\Delta_{00}$ for various ratios $\Gamma_N/T_{c0}$ in (a) Born limit; (b) unitarity limit; (c) combination of the two as employed in Ref.~\onlinecite{Lee-Hone:2017}.  (d) Residual scattering rate -2$ {\rm Im} \Sigma(\omega = 0)$ vs.\ $\Gamma_N/T_{c0}$  .}
\label{fig3}
\end{figure*}

For concreteness, in what follows we assume a separable pairing interaction $V_{{\bf k},{\bf k'}} = V_0 \Omega_{\bf k}\Omega_{\bf k'}$ in the  $d$-wave eigenfunction
$\Omega_\mathbf{k}$ defined in the first Brillouin zone of the two-dimensional CuO$_2$ planes,
\begin{equation}
\Omega_\mathbf{k} \propto \big(\cos(k_x a) - \cos(k_y a) \big)\;,
\end{equation}
where $a$ is the lattice spacing and $\Omega_\mathbf{k}$ is normalized such that $\langle \Omega_\mathbf{k}^2\rangle_\mathrm{FS} = 1$.

\subsection{Superfluid density}

The superfluid density, $\rho_s \equiv 1/\lambda^2$, and optical conductivity are closely related, with the spectral weight available to form the superfluid set by the normal state conductivity, via the Ferrell--Glover--Tinkham sum rule,
\begin{equation}
\int_0^{\omega_c} \sigma_1(\omega) \D \omega = \tfrac{\pi}{2} \epsilon_0 \omega_p^2\;.
\end{equation}
Here $\omega_p$ is the in-plane plasma frequency of the conduction electrons, with the cutoff $\omega_c$ chosen to capture the Drude weight of the free carriers, leading to 
\begin{equation}
\omega_p^2 = \frac{e^2 N_0}{\epsilon_0} \langle v_{F,x}^2\rangle_\mathrm{FS}\;.
\end{equation}
In the absence of disorder, the entire Drude weight would condense to give a clean-limit zero-temperature superfluid density
\begin{equation}
\rho_{s00} \equiv \frac{1}{\lambda_{00}^2} = \mu_0 \epsilon_0 \omega_p^2\;.
\end{equation}
Interaction effects reduce the plasma frequency below its bare value, and will be discussed later in the context of the cuprates.

Expressions for the penetration depth or superfluid density of a $d$-wave superconductor in the presence of disorder have been given in many places, and were specifically reviewed in Ref.~\onlinecite{Lee-Hone:2017}. We assume, as in most of these works, that nonmagnetic scatterers are pointlike so that impurity vertex corrections to the current--current correlation function 
vanish.\cite{Hirschfeld:1988}  Many references use a formulation that explicitly or implicitly assumes a circular Fermi surface for a quasi-2D system like the cuprates.  Here we work with a tight-binding model appropriate for overdoped \lsco,\cite{Yoshida:2006hw} which was shown in Ref.~\onlinecite{Lee-Hone:2017} to be crucial to understanding the $T$ dependence of the superfluid density.

Within our model, where we linearize the electronic structure near the Fermi surface, the finite temperature superfluid density in the presence of disorder is given by
\begin{equation}
\rho_s(T) = \mu_0 e^2 2 \pi T N_0\sum_{\omega_n > 0}\left\langle {v}_{F,x}^2\frac{\Delta_\mathbf{k}^2}{(\tilde \omega_n^2 + \Delta_\mathbf{k}^2 )^\frac{3}{2}} \right\rangle_{\mathrm{FS}}\;.
\label{superfluiddensity}
\end{equation}
At $T=0$, disorder suppresses $\rho_s$ from the clean-limit value $\rho_{s00}$.  In Fig.~\ref{fig1}, we have reproduced the results of Ref.~\onlinecite{Lee-Hone:2017} based on Eqs.~(\ref{unitarityBorn}) and (\ref{superfluiddensity}), and shown that the temperature and doping dependence compares semiquantitatively with the experimental results of Ref.~\onlinecite{Bozovic:2016ei}.

\subsection{Optical conductivity}

We now proceed to calculate the real (dissipative) part of the conductivity  $\sigma_1(\Omega)$ within the same framework.  Again, this has been done in many places earlier,\cite{Hirschfeld:1994,Graf:1995,Quinlan:1996} so we provide only the final expressions, which agree, e.g., with Refs.~\onlinecite{Hirschfeld:1994,Quinlan:1996}.  We  find
\begin{eqnarray}
 \sigma_1( \Omega ) &=& -\frac{N_0 e^2}{2 \Omega} \int_{-\infty}^{\infty} \mathrm{d}\omega \left[ f(\omega) -f(\omega+\Omega) \right]\times
\nonumber \\ &&\left\langle \, v_{F,x}^2 \,\mathrm{Re}\left\{ A_{++}-A_{+-}\right\}\right\rangle_\mathrm{FS}\;,
 \label{eq:conductivity}
\end{eqnarray}
where
\begin{eqnarray}
A_{+\pm} &=& \frac{\Delta_{\bf k}^2+\tilde\omega_+ \tilde\omega'_\pm + Q_+ Q' _\pm}{ Q_+ Q '_\pm \left(  Q_+ + Q '_\pm\right)} \\
\omega_\pm &=& \omega \pm i \eta \\
\omega'_\pm &=& \omega+\Omega \pm i \eta \\
Q_\pm &=& \sqrt{\Delta_{\bf k}^2-\tilde \omega^2_\pm} \\
 Q'_\pm &=& \sqrt{\Delta_{\bf k}^2-\tilde \omega'^2_\pm}\;,
\end{eqnarray}
and the renormalized real axis frequencies are $\tilde \omega_\pm (\omega) =\tilde\omega_n (i\omega_n\rightarrow \omega \pm i \eta)$. 
 Here the branch cut for the complex square root function is the negative real axis.
These equations, taken together with the definitions of the renormalized frequencies (\ref{tmatrix}), are sufficient to calculate the optical conductivity at all frequencies and temperatures.   In the normal state limit $\Delta_{\bf k}\rightarrow 0$, (\ref{eq:conductivity}) reduces to the Drude conductivity 
 \begin{equation}
\sigma_{1N}(\Omega) = \sigma_{N0}\left(\frac{4\Gamma_N^2}{\Omega^2+4 \Gamma_N^2}\right)\;,
\end{equation}
where $\sigma_{N0}$ is the DC conductivity  $e^2 N_0 \langle v_{F,x}^2 \rangle_\mathrm{FS}/(2\Gamma_N)$.
\vskip .2cm

\section{Results}
\subsection{Qualitative considerations}
            
It is essential to appreciate the unusual and sometimes counterintuitive effects of the different energy dependences of the impurity scattering rate of the $d$-wave superconductor in the Born and unitarity limits.  In Fig.~\ref{fig2}(a)-(c), we illustrate how the density of states varies with energy in the two limiting cases, as well as a special mixture of the two that was found in Ref.~\onlinecite{Lee-Hone:2017} to fit the superfluid density well.

\begin{figure*}[t]			
\includegraphics[width=1\textwidth]{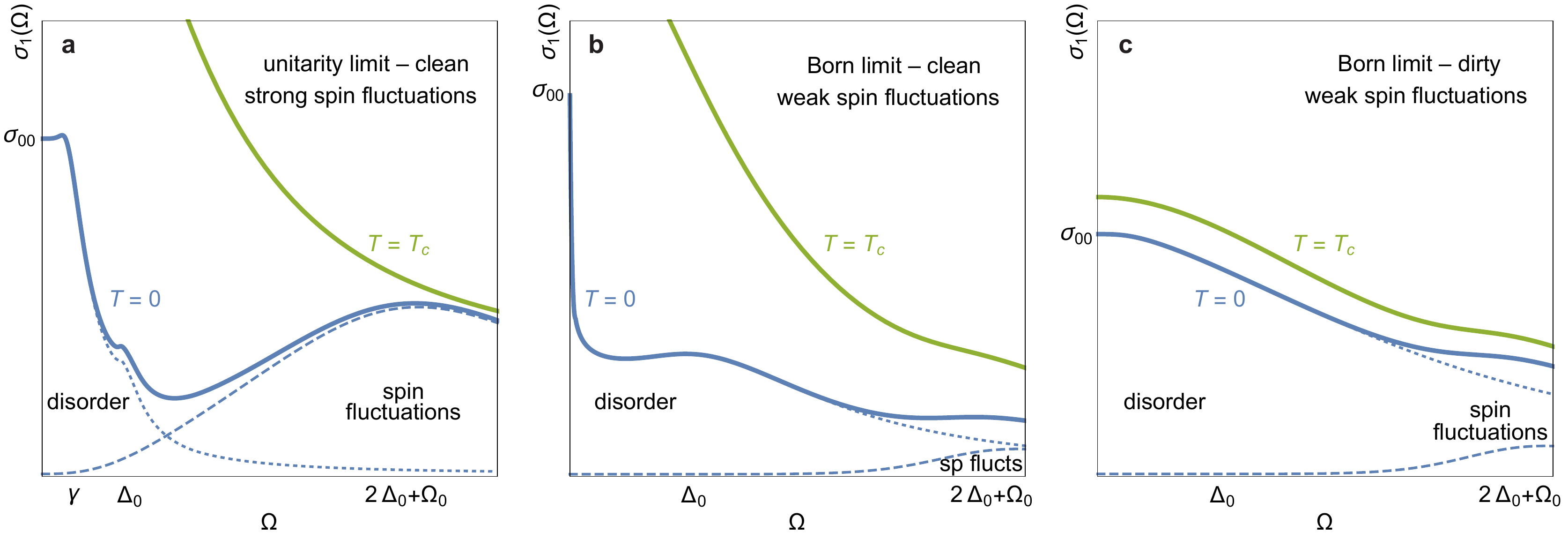}
			\caption{Schematic of observed conductivity spectrum at low and high temperatures.  (a) Unitarity limit disorder scattering plus strong spin fluctuation scattering, as proposed for optimally doped cuprates;\cite{Quinlan:1996} (b) Born limit disorder scattering in clean limit, assuming weak spin fluctuation scattering; (c) Born limit disorder scattering in dirty limit, assuming weak spin fluctuation scattering. Note the absence of a spectroscopic gap signature in the dirty limit.}
\label{fig4}
\end{figure*}  

    It is well-known that the residual density of states  $N(\omega\rightarrow 0)$ scales as $\gamma\sim \sqrt{\Gamma_N \Delta_0}$ in the unitarity limit, creating a plateau in $N(\omega)$ over a range of energies $\gamma$ (the ``impurity band"), at the expense of the coherence peak.  Relatively small scattering rates $\Gamma_N$ are sufficient to strongly modify these low-energy states.      
            In the clean Born limit, on the other hand, the corresponding density of states $N(0)$ is also finite, but scales with a $\gamma$ that is exponentially small in $\Delta_0/\Gamma_N$.\cite{Hirschfeld:1994}  Since a scattering resonance is never produced in this limit, all states are modified equally weakly, such that for a  range of small disorder negligible effects are seen on 1-particle spectral quantities.  In Fig.~\ref{fig2}(d), we compare the dependence of the residual DOS $N(0)$ on $\Gamma_N$, proportional to the impurity concentration.   From  Figs.~\ref{fig2}(a) and (d), it is clear that in the Born limit the scattering rate must reach a very large fraction of $T_{c0}$, i.e., a very large fraction of the critical rate to destroy superconductivity altogether, before the density of states is substantially modified over any significant range of energies.
            
            To study the optical conductivity, it is even more important to understand the differences in the energy dependences of the scattering rate in the Born and unitarity limits, shown in Fig.~\ref{fig3}(a)-(b).  It is easy to see from Eq.~(\ref{tmatrix}) that in the unitarity limit $c=0$, the scattering rate $1/\tau \equiv 2\,{\rm Im}\,\tilde \omega_+$ varies inversely proportional to $N(\omega)$ with the exception of the impurity band region, where the $1/\omega$ divergence is cut off by self-consistency.   On the other hand, $1/\tau$ is proportional to $N(\omega)$ in the  Born limit, again cut off at the lowest energies, but in a manner hardly visible in the clean limit.  In the dirty Born limit, the energy dependence of $1/\tau$ becomes generally smeared out over the entire energy range, a fact that will be important for our analysis below.

            The superfluid density  (\ref{superfluiddensity}) reflects the density of states directly.  In particular, the nonzero residual density of states $N(0)$ may be shown to lead directly to a $T^2$ term in the penetration depth \cite{Gross:1986} {\it  at sufficiently low $T$}.  This is confined to a temperature region of order $T^\ast\sim\gamma$ in the unitarity limit, and if $T^\ast\ll\Delta_0$ it is possible to observe a crossover linear-$T$ region as well.\cite{Hirschfeld:1993cka}  It is sometimes assumed that the same statement may be applied in the Born limit, with the only difference being the much smaller $\gamma$. However, as is clear from Fig.~\ref{fig2}(a), in the Born limit there is no such separation of energy scales, leading to a quasi-linear $T$ behavior in the penetration depth over the whole range, or, for sufficient disorder, a $T^2$ behavior over the whole range.\cite{Kogan:2013kf,Lee-Hone:2017}  As emphasized in Ref.~\onlinecite{Lee-Hone:2017}, the regime of  quasi-linear $T$ behavior in the intermediate $T$ regime extends to rather high disorder values for realistic band structure models appropriate for  underdoped \lsco.

From these considerations exhibited in Figs.~\ref{fig2} and \ref{fig3}, it is possible to sketch the expected behavior in the conductivity.  There are two additional factors that influence the peculiar form of the spectrum in a cuprate $d$-wave superconductor, however.  The first is the existence of a unique zero-frequency, zero temperature limit of $\sigma_1(\Omega)$, the so-called ``universal'' $d$-wave conductivity  $\sigma_{00}\equiv e^2 N_0 v_F^2 \hbar/(2 \pi \Delta_0)$, 
which is, unlike a normal quasiparticle conductivity, finite in the limit of vanishing disorder.\cite{Lee:1993}  The expression for $\sigma_{00}$ is not strictly independent of disorder, since it depends on the gap magnitude which is itself suppressed within the present theory, such that $\sigma_{00}$ is expected to increase slightly with addition of impurities.  Furthermore, it is enhanced by impurity vertex corrections if the impurity potential has nonzero range,\cite{Durst:2000} but these effects vanish  in our model with pointlike scatterers.

\begin{figure*}[t]		
\includegraphics[width= 0.98 \linewidth]{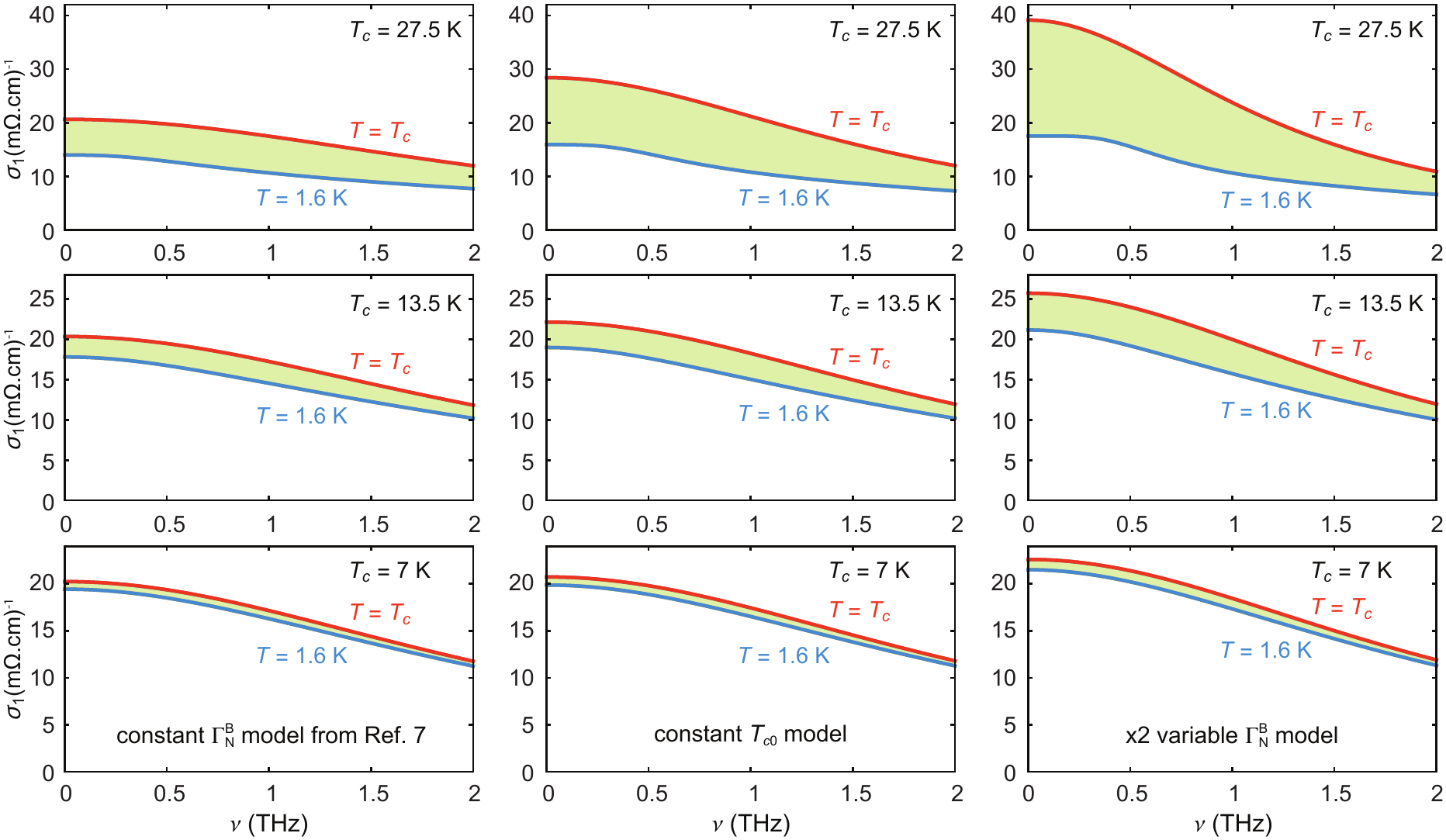}         			
\caption{Optical conductivity $\sigma_1(\nu)$ of overdoped \lsco\ calculated from Eq.~\ref{eq:conductivity}, to be compared with \mbox{Figs.~2a--c} of Ref.~\onlinecite{Mahmood:2017}.  Each column shows normal state ($T = T_c$, red) and low temperature ($T = 1.6$~K, blue) results for three different dopings, corresponding to critical temperatures $T_c = 27.5$, 13.5 and 7~K. The shaded areas in between show the spectral weight that condenses at low temperature to form the superfluid.  The calculated conductivities include a prefactor of 0.3 to accommodate the experimentally observed renormalization of plasma frequency, as shown in Fig.~\ref{fig7}.
The unitarity limit scattering parameter is fixed at \mbox{$\Gamma_N^{U}/\pi= 1$~K} in all cases.  Three different scenarios are presented for the doping dependence of the Born scatterers: (first column) constant $\Gamma_N^{B}/\pi= 17$~K, as in Ref.~\onlinecite{Lee-Hone:2017}; (second column) $\Gamma_N^{B}(p)$ adjusted such that $T_{c0}$ is doping independent; and (third column) $\Gamma_N^{B}(p)$ increasing linearly in doping by a factor of two on passing from $T_c = 27.5$~K to $T_c = 7$~K. Parameters for the three disorder models are plotted in Fig.~\ref{fig8}, and have been formulated to converge on the strongly overdoped side.}
\label{fig5}
\end{figure*} 

The second effect that must be accounted for is inelastic scattering by spin fluctuations.  The relaxation time of nodal quasiparticles that dominate the conductivity at low frequencies may be shown to vary as $1/\tau_\mathrm{sf}\sim \omega^3$ at $T=0$ for $\omega\ll 3\Delta_0$.\cite{Quinlan:1994}  This behavior of the relaxation time leads to a scenario for the shape of the conductivity spectrum seen in optimally doped cuprates, as discussed in Ref.~\onlinecite{Quinlan:1996}.  The low energy behavior is dominated by disorder physics in the form of a relatively narrow residual Drude component tied at $T=0$ to $\sigma_{00}$, while the higher energy physics (``mid-infrared component'') arises from the 
crossover between these low- and high frequency inelastic scattering regimes, and is found to peak at 4$\Delta_0$  in weak coupling theory,\cite{Orenstein:1990} or more generally at $2\Delta_0+\Omega_0$, where $\Omega_0$ is the frequency of the pairing boson.  This picture is sketched in Fig.~\ref{fig4}a, and leads to the familiar \mbox{low-$T$} conductivity spectrum of optimally doped cuprates, where spectral weight is lost at finite frequencies relative to the normal state in an intermediate frequency region between $\gamma$ and $2\Delta_0+\Omega_0$.  It should be noted that there are other proposals for the origin of the mid-infrared peak, discussed in Ref.~\onlinecite{BasovTimusk2005}.

The implications of weak scattering for the conductivity spectrum within this scenario have not been systematically explored.  In Figs.~\ref{fig4}(b) and (c), we show what is anticipated from the discussion of Figs.~\ref{fig2} and \ref{fig3} for disorder scattering in the Born limit. In the clean case, the conductivity must tend to $\sigma_{00}$ at low $T$, but the effect is essentially invisible due to the exponentially small residual scattering $\gamma$.   The disorder-limited  conductivity is then very small and flat out to high frequencies.\cite{Graf:1995}  We have depicted a situation where spin fluctuations scattering is weak, anticipating a loss of low frequency spin fluctuations in the overdoped cuprates under consideration.  Note in this limit one expects the spectral weight loss between $T=T_c$ and $T=0$ to be frequency dependent, although there is no well-defined region where it occurs, as in the unitarity scattering case.  In the dirty Born limit, on the other hand, the spectrum is completely dominated by disorder.  While it must still approach the value $\sigma_{00}$ as $T,\Omega \rightarrow 0$, in absolute units this value is larger due to the suppression of $\Delta_0$.  There is no well-defined residual Drude component, as all states over a wide energy range are broadened (see Fig.~\ref{fig2}b).  Because the zero-temperature superfluid density is considered to be strongly suppressed in this example, the $T=0$ conductivity should closely
follow the normal state spectrum.  Again, we have assumed in Fig.~\ref{fig3}c that spin fluctuations are weak, and may not play a major role.

\subsection{Disorder-limited optical conductivity}
We now present evaluations of Eq.~(\ref{eq:conductivity})  to illustrate explicitly the importance of the impurity phase shift to the conductivity spectrum.  Our specific goal is to compare the predictions of the disorder-based model elaborated in Ref.~\onlinecite{Lee-Hone:2017}, and show that it explains semiquantitatively the results of Mahmood et al.\ for the THz conductivity of overdoped \lsco.  In Fig.~\ref{fig5} we plot the optical conductivity in the normal state and at low temperatures, which should be compared with the experimental data in \mbox{Figs.~2a--c} of Ref.~\onlinecite{Mahmood:2017}. Each column shows results for three different dopings, corresponding to critical temperatures $T_c=27.5,13.5,$ and 7~K,   In the first column the impurity parameters and $T_{c0}$ are identical to those used in Ref.~\onlinecite{Lee-Hone:2017}.  In all cases the normal state curve has a pure Drude form with physical scattering rate $2\Gamma_N$.  In the dirty Born limit represented here (with small unitarity limit corrections that affect only the very low frequency range),  the superconducting state conductivity spectrum is also surprisingly Drude-like.  That spectral weight has indeed been lost and condensed into a zero-energy $\delta$-function is clear in all cases from the area between the normal state and low temperature curves, but there is no sign of a well-defined low-energy region where scattering states have been removed, as exhibited, e.g.,  in the unitarity limit scenario in Fig.~\ref{fig4}(a).  As doping increases, $T_{c0}$ decreases and the effect of disorder becomes more pronounced, shifting more weight from the condensate to the finite-frequency quasiparticle spectrum.   Despite the simplicity of the disorder-based model considered in Ref.~\onlinecite{Lee-Hone:2017}, the calculated conductivity shows semiquantitatively the same behavior as the THz experiments.  The only essential difference is the decrease of the magnitude of the conductivity as the system is overdoped, not captured here because a fixed disorder strength has been used for all dopings.\cite{Lee-Hone:2017}  Note as well that there is no signature of the superconducting gap, in either the model or the experiment, despite the fact that the frequency range covered (2~THz) corresponds to 50~K in temperature units.  This is a particular consequence of the \mbox{$d$-wave} order parameter, for which disorder (especially in the dirty Born limit) rapidly smooths out sharp gap features in both the density of states and conductivity, as shown in Figs.~\ref{fig2} and \ref{fig4}, respectively.

\begin{figure}[t]
\includegraphics[width=0.8\linewidth]{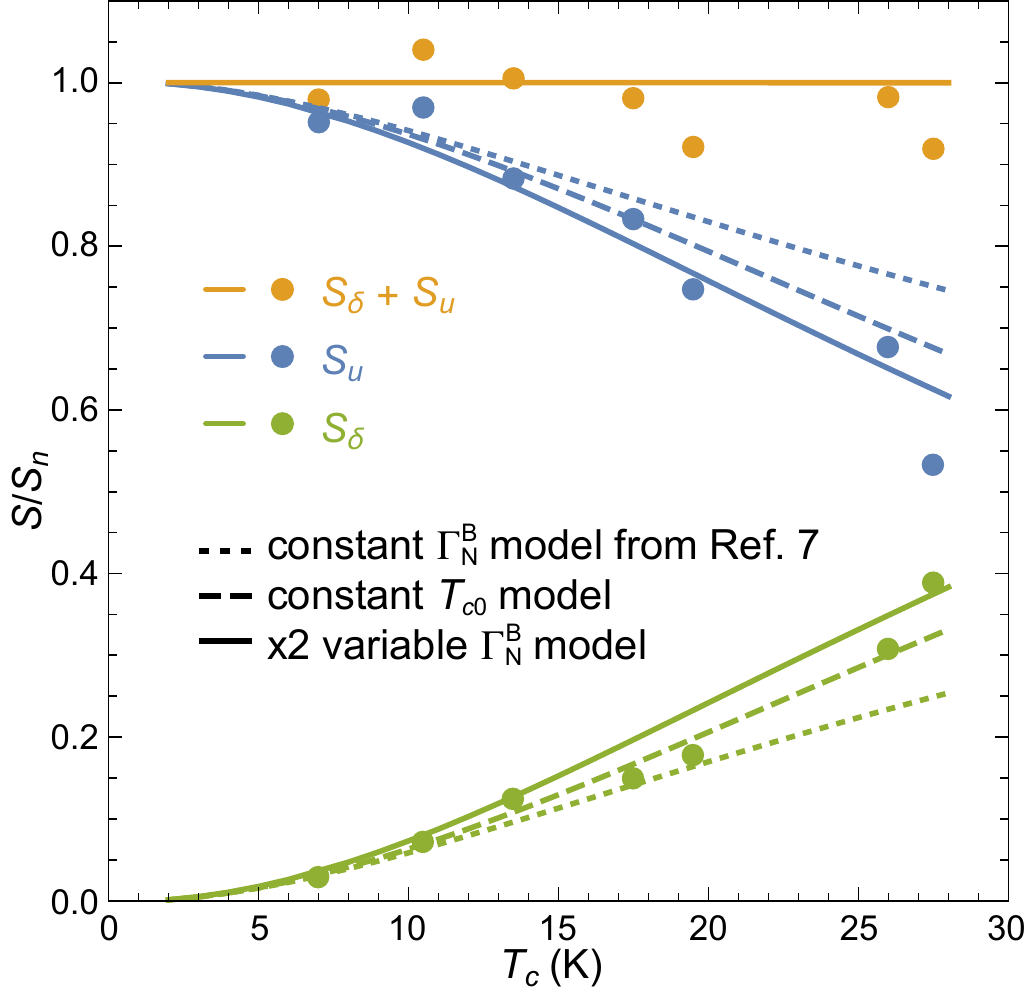}
\caption{Dimensionless optical spectral weights calculated at $T = 1.6$~K using the doping-dependent tight binding parameters discussed in Ref.~\onlinecite{Lee-Hone:2017} and the three disorder scenarios considered in Fig.~\ref{fig5}, compared with experiment. $S_u$ is uncondensed spectral weight from finite-frequency $\sigma_1(\Omega)$; $S_\delta$ is the condensed spectral weight in the superfluid; $S_n$ is the normal state spectral weight. Solid circles: experimental data from Ref.~\onlinecite{Mahmood:2017} at the same temperature, determined by Drude fits to $\sigma_1(\Omega)$ over the frequency range 0.3 to 1.7~THz.}
\label{fig6} 	
\end{figure}

The doping dependence of the  uncondensed dimensionless spectral weight loss can now be easily calculated by integrating  $S_u=2/(\pi \epsilon_0 \omega_p^2)\int_0^\infty  \sigma_1(\omega) \D \omega$. At the same time, the condensed spectral weight is obtained from the superfluid density via $S_\delta=\rho_s(T)/(\mu_0\epsilon_0 \omega_p^2)$, with the Ferrell--Glover--Tinkham sum rule implying that $S_u+S_\delta=1$ at any temperature. In Fig.~\ref{fig6}, we plot the components of the spectral weight as a function of doping at $T=1.6$~K, illustrating how the condensed spectral weight indeed vanishes with doping, even as the total carrier density ($\sim\omega_p^2)$ in the model  remains finite, a signature of the disorder-dominated regime.   In the same figure are plotted the experimental data obtained by Mahmood et al.\ by fitting a Drude form over the frequency range range from 0.3 to 1.7~THz.  It is easy to see that the model of fixed disorder scattering rate $\Gamma_N$ considered in Ref.~\onlinecite{Lee-Hone:2017} (dotted lines) provides a very reasonable fit to the data except for the samples closest to optimal doping, which we discuss further below.  

\section{Discussion}

\begin{figure}[t]
\includegraphics[width=0.8\linewidth]{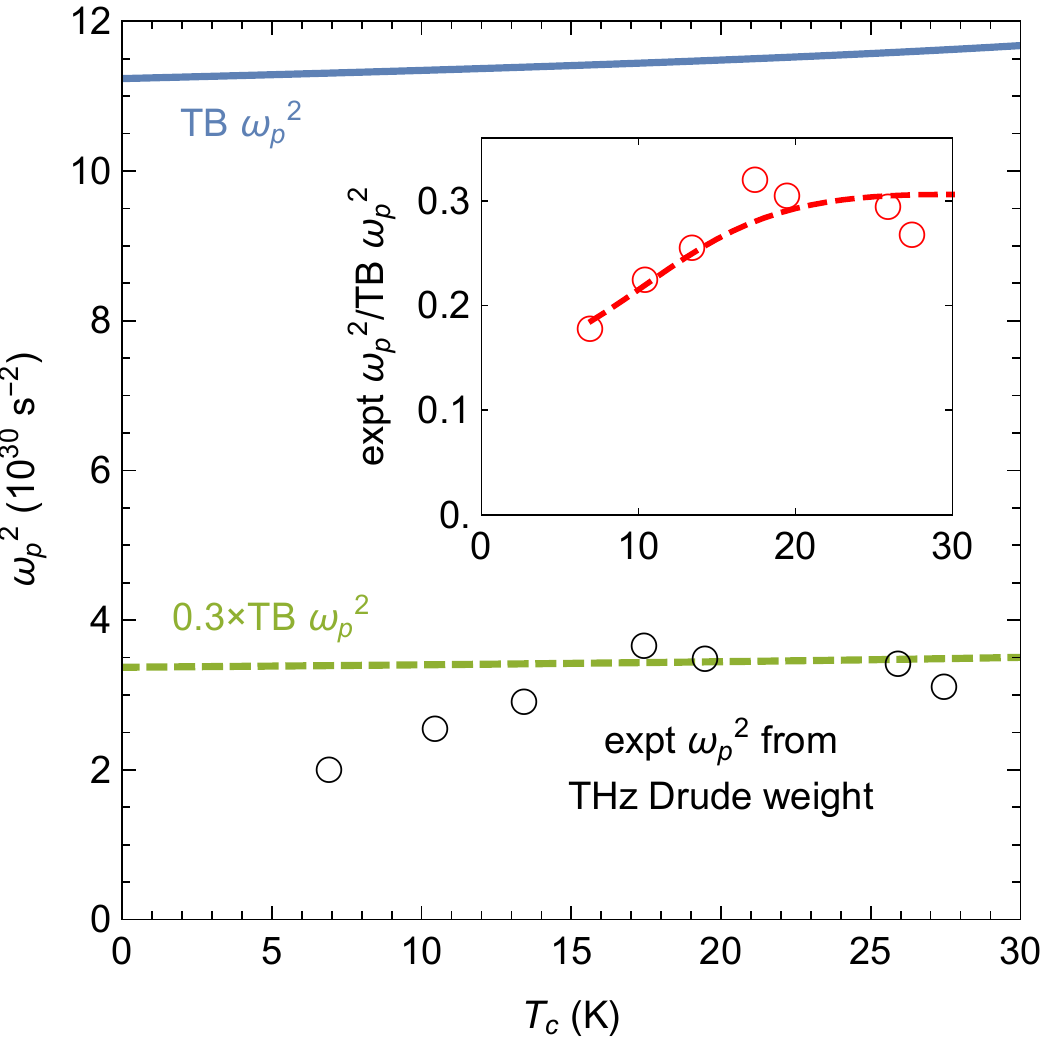}
\caption{Plasma frequency of \lsco.   Solid curve (TB~$\omega_p^2$) shows the square of the plasma frequency calculated from the doping-dependent tight-binding dispersion from Ref.~\onlinecite{Lee-Hone:2017}.  Open circles are the experimental $\omega_p^2$ inferred from  fitted normal-state Drude weights as reported in the THz conductivity study of Ref.~\onlinecite{Mahmood:2017}.   Inset: ratio of experimental $\omega_p^2$ to tight-binding $\omega_p^2$.}
\label{fig7} 	
\end{figure}

\subsection{Renormalized plasma frequency}

It is well known in interacting electron systems that plasma frequency and Drude weight are reduced below the bare values calculated from band theory.\cite{ScalapinoWhiteZhang:1992,Padilla:2005ir,Basov:2011ht}  This arises both from many-body effects that locally flatten the dispersion near the Fermi surface,\cite{Lanzara:2001kp,Zhou:2003,Li:2018gf} simultaneously transferring  spectral weight to higher frequencies;\cite{Basov:2011ht} and from Fermi-liquid effects,\cite{Gross:1986,Mesot:1999ir,Durst:2000,Paramekanti:2002cr} in which residual quasiparticle interactions induce a backflow that partially cancels the current response.
These effects can be accounted for by employing a renormalized plasma frequency.  To gauge the magnitude of the renormalization in overdoped \lsco, Fig.~\ref{fig7} plots $\omega_p^2$ calculated directly from the ARPES-derived tight-binding energy dispersion 
(TB $\omega_p^2$) and compares it to fitted Drude weights obtained from the THz experiments (expt $\omega_p^2)$. The calculated $\omega_p^2$ is larger than the experimental value by a factor of 3 to 4, consistent with previous work on the \lsco\ system.\cite{Padilla:2005ir} Indeed, the authors of the ARPES study from which the \lsco\ tight-binding band structure was obtained\cite{Yoshida:2006hw} point out that their fitted tight-binding dispersion does not capture the electron--phonon kink structure in $\epsilon_\mathbf{k}$ that appears near the Fermi energy.\cite{Yoshida:2007}  The inset of Fig.~\ref{fig7} plots the overall renormalization of the Drude weight, given by the ratio of the experimental to tight-binding $\omega_p^2$.  While some doping dependence is apparent, for the purposes of our calculation we  simply assume the renormalization takes a doping independent value of 0.3, and have applied this as a prefactor to the conductivities in Fig.~\ref{fig5}.  In contrast, for the dimensionless spectral weights plotted in Fig.~\ref{fig6} the renormalization of plasma frequency cancels out, making this measurement a particularly robust context in which to compare theory and experiment. 

\subsection{Clean limit critical temperature parameter $T_{c0}$}
In this work, for simplicity, we have considered an idealized, weak-coupling BCS theory.  The model corresponds to a situation in which the characteristic frequency of the exchange bosons responsible for Cooper pairing, $\omega_0$, is much greater than the superconducting transition temperature.  This leads to a pairing interaction that is roughly independent of frequency, allowing the combined effects of coupling strength and boson frequency to be captured by a single parameter, the notional clean-limit transition temperature $T_{c0}$.  Although beyond the scope of this paper, a more realistic model, e.g., based on spin fluctuations, would result in a boson spectrum spread over a wide range of frequencies.  As discussed in Ref.~\onlinecite{Millis:1988}, low frequency boson spectral weight, at $\omega \le T$, gives rise to inelastic electron--boson scattering, leading to significant pair breaking in unconventional superconductors.  Inelastic scattering is known to be particularly strong in optimally doped cuprates and so in this regime we expect the $T_{c0}$ parameter of the weak-coupling theory to substantially overestimate the $T_c$ of the hypothetical disorder-free reference material that might be obtained, say, by gating.  

\begin{figure}[t]
\includegraphics[width=0.85\linewidth]{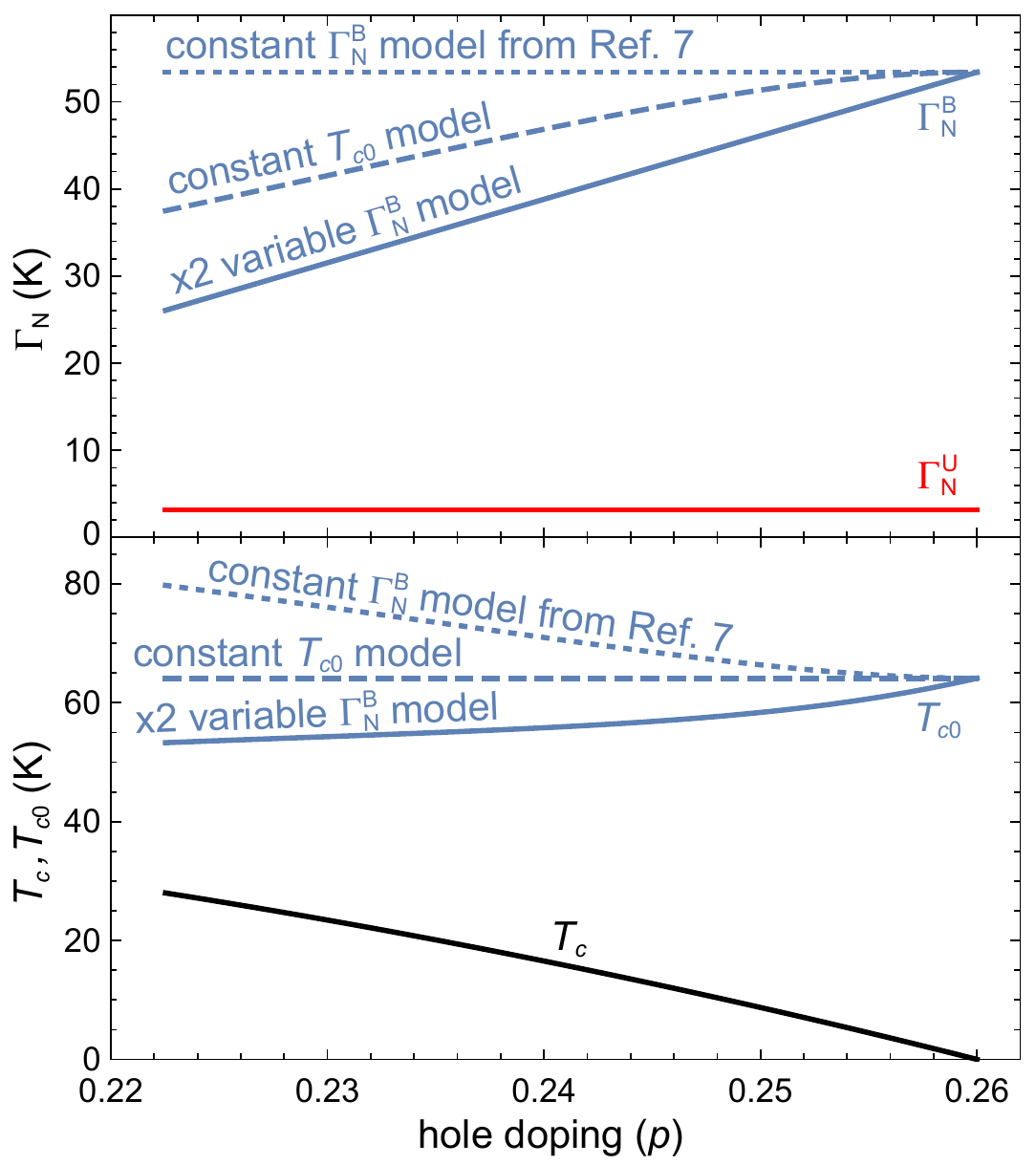}
\caption{
(upper panel) The three disorder scenarios considered in Fig.~\ref{fig5}: constant $\Gamma_N^{B}$, as in Ref.~\onlinecite{Lee-Hone:2017}; $\Gamma_N^{B}$ adjusted to give constant $T_{c0}$; and $\Gamma_N^{B}$ varying by a factor of 2 over the doping range of interest. In all cases the unitarity limit scattering parameter, $\Gamma_N^{U}$, is fixed at the same doping independent value. (lower panel) Superconducting transition temperature, $T_c$, as a function of doping, along with the implied doping dependence of $T_{c0}$ in the three scenarios.}
\label{fig8} 	
\end{figure}

There is one region of the phase diagram where the parameters of the weak-coupling model can be checked relatively directly against experiment: near the overdoped critical point at which $T_c \to 0$.  In the zero-temperature limit, inelastic scattering is irrelevant and pair-breaking can be attributed entirely to disorder.  The measured transport scattering rate $\tau^{-1}_\mathrm{tr}$ then places a lower bound on the single-particle elastic scattering rate, $\tau^{-1}_\mathrm{sp} =  2\mathrm{Im}\{\tilde \omega_+\}$, where in the normal state $ \mathrm{Im}\{\tilde \omega_+\} \equiv \Gamma_N$.  In the THz conductivity study of Mahmood et al.,\cite{Mahmood:2017} the most overdoped sample ($T_c = 7$~K) has a normal-state spectrum with Drude width 1.75~THz, corresponding to \mbox{$\Gamma_\mathrm{tr} = 84$~K} in temperature units, placing a lower bound on the normal-state scattering rate such that $\Gamma_N \ge 42$~K.  Equivalently, we can obtain a lower bound for the clean-limit transition temperature by taking the Abrikosov--Gor'kov result for the critical scattering rate of a weak-coupling \mbox{$d$-wave} superconductor, $\Gamma_c = 0.881 T_{c0}$, from which we infer $T_{c0} \ge 48$~K.  Against these experimental bounds, the values assumed in our weak-coupling model, $\Gamma_N = 56.5$~K and $T_{c0}(p \to p_c) = 64$~K, are not unreasonable.

\subsection{Disorder dependence of $T_{c0}$ and $\Gamma_N$  }

In Fig.~\ref{fig6}, the small discrepancies for the more optimally doped samples between  experimental values of the uncondensed spectral weight compared to the solid line representing the present theory    may reflect one of two things: a) these samples are closer to the clean limit, meaning that the fit performed on the experimental data over a limited intermediate frequency range misses   some weight at the lowest frequencies by extrapolating the Drude form to DC (see, e.g., Fig.~\ref{fig4}(b)); or b), we have until now neglected  the doping dependence of the scattering rate (\`a  la Ref.~\onlinecite{Lee-Hone:2017}). The actual sources of disorder and their scattering potentials are not completely characterized,  but it seems reasonable to assume that the small number of unitarity scatterers are Cu vacancies, and that the weak out of plane disorder is a combination of Sr dopants, whose number is reasonably well known, and O vacancies, whose number is poorly known.  To give some idea  of the sensitivity of our results to differing assumptions about these sources, we fix the 
number of unitarity scatterers but consider two additional models where the Born limit contribution, $\Gamma_N^{B}$, varies with doping but is normalized to give the same result as the Ref.~\onlinecite{Lee-Hone:2017} model at $T_{c}=0$.  
In one, $\Gamma_N^{B}(p)$ increases linearly with doping by a factor of two over the relevant range, in rough correspondence to what is observed in the terahertz experiments;\cite{Mahmood:2017} in the second, we vary $\Gamma_N^{B}(p)$ so as to give a constant $T_{c0}$  vs.\ hole doping.  The  normal state scattering rates for both
the Born and unitarity scatterers, along with $T_c$ and $T_{c0}$, are plotted vs.\ hole doping in Fig.~\ref{fig8}.   

The values of $T_{c0}$ required to fit the experimental data in the three models of doping dependence are all substantially larger than $T_c$ itself, so this qualitative feature of the weak-coupling theory remains.  It is interesting to note, however, that the fit to the spectral weight loss in Fig.~\ref{fig6}  improves markedly in the near-optimally doped samples when one includes the doping dependence of scattering.  In fact, the lower scattering rate closer to 
optimal doping suggests that cleaner samples may have a weak  low $\Omega$ residual Drude upturn below the experimental observation window.

\section{Conclusions}

We have shown that a model of ``dirty $d$-wave" superconductivity used to fit superfluid density data by Bozovic et al.\cite{Bozovic:2016ei} on high quality overdoped \lsco\ films is also capable of fitting most features of the terahertz conductivity measured on similar samples\cite{Mahmood:2017} {\it with identical parameters.}  This is surprising given that scattering rates deduced for these films from Drude fits in the normal state appeared to be
much larger than $T_c$.   The explanation for the excellent description of two independent aspects of the electromagnetic response requires two assumptions.  First, we have postulated  that the intrinsic critical temperature in the absence of disorder is substantially higher than $T_c$ itself over the entire optimally doped range, and discussed why this may be reasonable.  This ansatz is indirectly supported, in fact, by the recent observation of the persistence of strong magnetic fluctuations to high doping in RIXS.\cite{Dean:2013}   The effective ``pure $T_{c0}$'' if one goes beyond BCS and includes  inelastic scattering in the theory may be substantially lower, but we do not expect that these effects will alter our qualitative conclusion that disorder is playing an essential role in current measurements of the vanishing of $T_c$ on the overdoped side of the cuprate phase diagram.  

The second crucial aspect of the theory that allows for  strong suppression of superfluid density, linear dependence of superfluid density on $T_c$, and Drude-like conductivity spectra at low temperatures,  is the unusual  role of weak scattering.  Here we have explored the effect  of Born scattering on the $d$-wave conductivity in some detail, and tried to exhibit some of the more  salient qualitative aspects.  A more complete theory of  the  conductivity spectrum and the overdoped phase diagram may require treatment of forward impurity scattering and a more proper treatment of the dynamics of the pairing bosons.  The semiquantitative success of the present theory suggests strongly, however, that exotic physics beyond Fermi liquid and Eliashberg  theory is probably not  required to understand the data.

 We emphasize that the possibility of a ``dirty $d$-wave'' explanation  for the apparently surprising results of Bozovic et al.\cite{Bozovic:2016ei} and Mahmood et al.\cite{Mahmood:2017} does {\it not} imply that the overdoped state of the cuprates can be described by a weakly interacting Fermi gas with a $d$-wave pairing instability.  While the dirty $d$-wave results fit the superfluid density and conductivity data well, we have also shown that a large renormalization of the plasma frequency by many-body effects, consistent in magnitude with that reported by other authors for this system, is required.  Renormalizations of similar magnitude have been reported for other cuprates for some time.\cite{Liu:1999}  A reliable  theory of the cuprates, even on the overdoped side of the phase diagram, must therefore account simultaneously for these effects on a microscopic basis.   

\acknowledgments

We acknowledge useful discussions with N.~P.~Armitage, L.~Benfatto,  J.~S.~Dodge, M.~P.~Kennett, S.~A.~Kivelson, F.~Mahmood, J.~Orenstein, D.~J.~Scalapino, and C.~M.~Varma.   We gratefully acknowledge financial support from the Natural Science and Engineering Research Council of Canada, the Canadian Institute for Advanced Research, and the Canadian Foundation for Innovation.  PJH was supported by NSF-DMR-1407502.


%

\end{document}